\documentclass[10pt,preprint]{aastex}
\usepackage{bm,natbib}
\usepackage{pdflscape}

\newcommand{\myemail}{a.juarez@vanderbilt.edu}
\newcommand{\Ha}{$\rm H\alpha$}
\newcommand{\Li}{Li{\sc \,i}}
\newcommand{\eg}{{\em e.g.},}

\slugcomment{Accepted in the Astrophysical Journal}
\shorttitle{Improved Determination of the LDB Age of Blanco~1}
\shortauthors{Juarez, Cargile, et al.}

\begin{document}
\title{An Improved Determination of the Lithium Depletion Boundary Age of Blanco~1 and a First Look on the Effects of Magnetic Activity}
\author{Aaron J.\ Juarez\altaffilmark{1,2}, Phillip A.\ Cargile\altaffilmark{2}, David J.\ James\altaffilmark{3,4}, Keivan G.\ Stassun\altaffilmark{1,2}}
\altaffiltext{1}{Department of Physics, Fisk University, Nashville, TN 37208, USA}
\altaffiltext{2}{Department of Physics and Astronomy, Vanderbilt University, Nashville, TN 37235, USA, \myemail}
\altaffiltext{3}{Cerro Tololo Inter-American Observatory, Casilla 603, La Serena, Chile}
\altaffiltext{4}{Qatar Environment and Energy Research Institute, Qatar Foundation, Tornado Tower, Floor 19, P.O. Box 5825, Doha, Qatar}

\begin{abstract}
The Lithium Depletion Boundary (LDB) is a robust method for accurately determining 
the ages of young clusters, but most pre-main-sequence models used to derive LDB ages do not include 
the effects of magnetic activity on stellar properties. In light of this, we present results 
from our spectroscopic study of the very low-mass members of the southern open cluster 
Blanco~1 using the Gemini-North telescope, program IDs: GN-2009B-Q-53 and GN-2010B-Q-96. 
We obtained GMOS spectra at intermediate resolution for cluster candidate members with 
$I$$\approx$13--20 mag. From our sample of 43 spectra, we find 14 probable cluster members 
by considering proximity to the cluster sequence in an $I/I-K\rm_s$ color-magnitude diagram, 
agreement with the cluster's systemic radial velocity, and magnetic activity as a youth 
indicator. We systematically analyze the \Ha\ and Li features and update the LDB age of
Blanco~1 to be $126^{+13}_{-14}$ Myr. Our new LDB age for Blanco 1 shows remarkable coevality
with the benchmark Pleiades open cluster. Using available empirical activity corrections, we
investigate the effects of magnetic activity on the LDB age of Blanco~1. Accounting for
activity, we infer a corrected LDB age of $114^{+9}_{-10}$ Myr.
This work demonstrates the importance of accounting for magnetic activity on LDB inferred stellar ages,
suggesting the need to re-investigate previous LDB age determinations.
\end{abstract}

\section{Introduction}

As low-mass stars ($\lesssim 1 M_\odot$) contract along the pre-main-sequence (PMS), their internal 
temperature rises. When the temperature of the stellar interior reaches $\sim$$2.5 \times 10^6$ K, 
lithium is destroyed by $^{7}$Li($p,\alpha)^{4}$He and $^{6}$Li($p,\alpha)^{3}$He proton capture 
reactions \citep[\eg][]{bodenheimer:1965}.
The elapsed time to reach Li-burning temperatures is a sensitive function of mass and thus, depends 
very sensitively on the luminosity \citep{bildsten:1997, ushomirsky:1998}. PMS low-mass stars are fully 
convective, so the mixing timescale is short, and since the temperature dependence of the nuclear 
reactions is steep, these stars rapidly deplete their Li content. For coeval stellar groups, like open 
clusters or moving groups, the determination of the luminosity at which stars transition from 
exhibiting Li in their atmospheres to being fully depleted provides a very precise age estimate. 
Moreover, the LDB technique is relatively model-insensitive, rendering similar ages to within 
$\pm$10\% \citep{burke:2004vf}, making it a highly robust method which can lead to the identification 
of missing input physics when compared with other age-dating methods. The LDB method originated with
\citet{basri:1996} who first applied it to the Pleiades cluster, leading to the discovery of the
first brown dwarf using the lithium test.

Ages of open clusters are traditionally determined by matching their Hertzsprung Russell diagrams (HRDs) 
to distance-dependent, theoretical stellar isochrones.  In particular, stars in and near the region 
close to the main-sequence turn-off (MSTO) are the most sensitive determinants of the age because 
they are evolving quickly.  MSTO fitting can be improved in precision through statistical techniques 
that take account of the usually small number of stars at the turn-off, but the method remains limited 
by several factors.  For very young clusters ($< 100$ Myr), the MSTO corresponds to the minimum number 
of objects for their initial mass function, and un-resolved, undetected binary/multiple systems can 
significantly affect MSTO ages.  At the same time, derived ages are highly dependent on the input 
models used and the physical constraints bounding them.  One major influence, for example, has been 
the inclusion of core mixing in intermediate-mass stars that led to ages being systematically 
increased by $\sim$50\% for clusters with ages less than 1--2 Gyr \citep{maeder:1974,naylor:2009}.
The LDB method offers a means 
to critically test the MSTO technique and it has several advantages. First, the physical processes 
involved in MSTO and LDB stars are completely different and so are independent.  Second, the fundamental 
physics underpinning the LDB method is much more simple and straight-forward to compute and calibrate 
than for hot, high-mass stars because stars lying close to the LDB point in young clusters are 
fully convective.  While the exteriors of very-low-mass stars ($< 0.2 M_\odot$) may host complex magnetic
activity phenomena that are challenging to understand, their interiors are fairly straightforward.
Third, the nature of open cluster mass functions implies that there are many more stars that can be
exploited to establish the LDB than there are at the MSTO \citep{soderblom:2010,soderblom:2013}.
However, one may obtain a relatively sparse data set due to efforts of removing field star contaminants.

The LDB method has limitations in its applicability. It can only be applied to very young 
clusters because of the rapidity of Li depletion. Furthermore, very-low-mass stars at the LDB are 
extremely faint, so only very nearby clusters are amenable to observation, usually by 8--10m class 
telescopes. Such stars are typically mid-M dwarfs in the cluster, and for physical and practical reasons, 
the LDB method is limited in its utility for ages in the range $20 <\tau< 200$ Myr.  Currently, 
seven other open clusters and two moving group associations have LDB age determinations, but only 
the Pleiades is similar in age \citep[125$\pm$8 Myr,][]{stauffer:1998aa} \citep[126$\pm$11 Myr,][]{burke:2004vf}
\citep[gyrochronology age of $134^{+9}_{-10}$ Myr,][]{cargile:2014} to Blanco~1.
As yet, no open cluster with an isochrone age $>$130 Myr has been investigated using the LDB method.

Blanco~1 is an open cluster whose near-solar composition, [Fe/H] $=+0.04\pm0.04$ \citep{ford:2005}, 
and age similarity to the Pleiades make it ideal for direct comparison and systematic characterization 
of age diagnostics. Previous age estimates for Blanco~1 suggest that it is a relatively young open cluster
\citep[100--150 Myr,][]{panagi:1997, moraux:2007}.
Blanco~1 is also considered nearby at a modest distance of 207 pc \citep{van-leeuwen:2009},
and lies at high Galactic latitude ($b=-79^\circ$). Initially, a subset of low-mass Blanco~1 
candidates were analyzed, and the LDB age was determined to be 132$\pm$24 Myr \citep{cargile:2010}. 
In this manuscript, we present additional Gemini-N spectra of Blanco 1 LDB candidates, and describe 
a consistent analysis for the full sample of Blanco 1 spectroscopic observations, which allows
us to further resolve the LDB location and derive a more precise LDB age for the cluster.

Despite the similarity of derived LDB ages among different PMS models, most models do not account 
for physical processes in an inclusive and realistic stellar environment, such as rotation and 
magnetism prevalent in low-mass star PMS evolution. Such omissions have the potential to affect 
the rate of Li depletion and thus, the age inferred from the LDB. It is well-established that 
magnetic activity can influence stellar parameters, particularly the radius ($R$) and effective 
temperature ($T_{\rm eff}$) of a star \citep{morales:2008}. One additional goal of this project 
is to quantify the extent that activity influences these stellar parameters, allowing us to 
correct for the age determinations based on the LDB technique. Using empirical relationships 
presented in \citet{stassun:2012}, we will account for the magnetic activity by effectively 
determining the properties of inactive Blanco~1 stars. In doing so, we enable a consistent LDB 
age determination from standard PMS models.

In Section~\ref{data}, we describe the data arising from our new medium-resolution spectroscopic 
campaign of additional Blanco 1 low-mass candidate members. In Section~\ref{analysis}, we present 
our analysis, emphasizing the \Ha\ and \Li\ features, which are important to the astrophysical 
interpretations for inferring the LDB age of Blanco~1. In Section~\ref{results}, we present the 
details of a clear methodology for establishing LDB boundaries as well as the derivation of the 
LDB age for our sample; we then present the empirical corrections for magnetic activity and derive 
a new LDB age based on the changes found in $T_{\rm eff}$ and $R$ for the stars which define the
LDB boundaries. We conclude the manuscript with a summary of our work in Section~\ref{summ}.

\section{Targets, Observations, and Data Reduction}\label{data}

A photometric catalog of the very-low-mass members of Blanco~1 was compiled by \citet{moraux:2007},
where they selected cluster candidates on the basis of their location in CMDs compared
to theoretical isochrones (100 and 150 Myr). \citeauthor{moraux:2007} furthermore took low-resolution
optical spectra for 17 of the brightest brown dwarf candidates and found \Ha\ in emission for 5 of them,
which is an initial indicator of youth. Their list of 15 probable members straddling the substellar
boundary provides us with an ideal sample for investigating the LDB of Blanco~1. These probable low-mass
cluster members have $I$$\approx$18--20, corresponding to the expected luminosity of Blanco 1 LDB, which
in a cluster of age $\sim$100 Myr at a distance of $\sim$200 pc should be $I$$\approx$19 
\citep{burke:2004vf,cargile:2010}.

For a subset of these objects, \citet{cargile:2010} have previously presented medium-resolution 
spectra of the \Li\,(6707.8\,\AA) region. In this work, we have obtained additional spectra using
the same instrument and setup as were employed for the Cargile et al. study. Both the previous and 
new spectra were obtained with the Gemini Multi-Object Spectrograph (GMOS) in queue schedule 
mode on the Gemini-North telescope \citep{hook:2004}, under program IDs GN-2009B-Q-53 and 
GN-2010B-Q-96. We used $1''$ slitlets to yield a two-pixel resolving power of $\simeq$4400 
over a spectral wavelength range of 5700--8000\AA\ and dispersion of 0.67\AA\ per pixel. 

Moreover, a recent optical survey performed using the SMARTS 1.0m telescope at CTIO provided 
additional candidates with $I$$\approx$13.0--17.5; these targets were identified as photometric 
candidate members from their location near to the cluster sequence in an optical color magnitude 
diagram (CMD, James et al. in prep).  Altogether, our sample contains 43 spectra (13 targets 
selected from Moraux et al., 30 from the SMARTS survey), from which, we find 14 high confidence 
members of the Blanco 1 cluster. In addition, we retain spectra of the radial velocity (RV) 
standard star GJ~905 (M6) from our initial GN-2009B-Q-53 program, which was observed and 
analysed in an identical manner to the Blanco~1 candidates.

All of our GMOS spectra are reduced using standard reduction routines in the IRAF\footnote{{\sc iraf} 
is distributed by the National Optical Astronomy Observatories, which are operated by the 
Association of Universities for Research in Astronomy, Inc., under cooperative agreement with 
the National Science Foundation.} Gemini-GMOS package, including bias removal, flat-fielding, 
aperture extraction, and wavelength calibration. Our spectral signal-to-noise ratios (SNR) per 
pixel ranged from approximately 10 to 500 for the faintest and brightest targets, respectively. 
RVs for each Blanco~1 target were measured using the \verb+fxcor+ procedure in IRAF by 
cross-correlating target GMOS spectra with the RV standard star, GJ~905. We performed 
cross-correlation in the wavelength region $\sim$6600--7000\AA, masking out regions rich 
in telluric features. Uncertainties on these RVs can be relatively large (up to $\sim$15 
km~s$^{-1}$), which is primarily due to the low SNR of the target spectra and the medium 
resolution of our observations.

\section{Analysis}\label{analysis}

We developed a spectral analysis code in Python to completely automate the analysis method in order 
to consistently derive equivalent widths and spectral indices. The spectral type of the object is 
determined via TiO and CaH spectral indices, whose methodology we describe in \S~\ref{spec}. 
Equivalent widths (EWs) of the \Ha\,(6562.8\,\AA) and \Li\,(6707.8\,\AA) spectral features are 
measured systematically, using established wavelength regions flanking both features to carry 
out the linear normalization procedure to the pseudo-continuum. Systematic EW measurement and 
error estimation for the \Ha\ line is discussed in \S~\ref{ha}. Cluster membership criteria 
are laid out in \S~\ref{mem}, where we identify high confidence cluster members. In 
\S~\ref{lithium}, we describe the measurement of the \Li\ feature, which we then place in 
the context of the curve of growth to derive lithium abundance, A(Li), in section~\ref{liab}, 
allowing us to provide insight as to the stage of Li depletion for our targets.
In section~\ref{activity}, we describe how we obtain $\log L_{\rm H\alpha}/L_{\rm bol}$ values, 
which are used to account for magnetic activity and determine their concomitant changes in 
$R$ and $T_{\rm eff}$. Observational and empirical data for the cluster members are summarized 
in Table~\ref{b1}, which include positions, photometric properties, spectral types, RVs, EW(\Ha), 
and EW(\Li) for each object. We include 1$\sigma$ errors for both the \Ha\ and \Li\ EW measurements 
for completeness, with 3$\sigma$ upper limits reported for Li non-detections.

\begin{landscape}
\begin{deluxetable}{lccccccccccc}
\tabletypesize{\footnotesize }
\tablewidth{0pt}
\tablecaption{Blanco 1 members ordered by intrinsic $I$ magnitude. \vspace{-4mm}}
\tablehead{
\colhead{Star Name\tablenotemark{a}} & \colhead{RA\tablenotemark{b}} & \colhead{Dec\tablenotemark{b}} &
\colhead{$I_0$\tablenotemark{c}} & \colhead{$(I-K_{\rm s})_0$\tablenotemark{c}} & \colhead{SpT\tablenotemark{d}} & \colhead{RV} &
\colhead{EW(\Ha)\tablenotemark{e}} & \colhead{$\log L_{\rm H\alpha}/L_{\rm bol}$\tablenotemark{f}} &
\colhead{EW(\Li)\tablenotemark{g}} & \colhead{A(Li)}\\
\phantom{yo!}&[HH:MM:SS]&[DD:MM:SS]&[mag]&[mag]& & [km s$^{-1}$]&[\AA]&  &[\AA]&}
\startdata
B1opt-{\bf 6335}    &  00:00:28.868 & -30:08:30.01  &$13.156\pm0.011$&$1.724\pm0.024$&   K5.5   &  $-10 \pm  8$  &   $-2.39 \pm 0.06$   &   $-3.6759$   &$0.087\,^{+0.087}_{-0.077}$ &$ 0.945\,^{+0.461}_{-0.592}$\\
B1opt-{\bf 18229}   &  00:01:39.768 & -30:04:38.24  &$13.315\pm0.016$&$1.703\pm0.029$&   K5.3   &  $ 15 \pm 12$  &   $-1.19 \pm 0.12$   &   $-3.9698$   &   $<0.249$               &$<-1.348$\\
B1opt-{\bf 2156}    &  00:07:40.790 & -30:05:56.58  &$14.45\pm0.030$&$2.09\pm0.042$&   M0.4   &  $ 7  \pm  6$  &   $-3.88 \pm 0.06$   &   $-3.6883$   &   $<0.177$  			   &$<1.593$\\
B1opt-{\bf 13328}   &  00:04:22.733 & -30:23:06.00  &$15.86\pm0.001$&$2.27\pm0.051$&   M3.8   &  $ 1  \pm  5$  &   $-7.89 \pm 0.04$   &   $-3.5062$   &   $<0.054$               &$<0.698$\\
CFHT-BL-{\bf 16}  &  00:01:28.438 & -30:06:06.95  &  18.30  &   2.85   &   M5.1   &  $ 4  \pm  6$  &   $-4.50 \pm 0.14$   &   $-4.1908$   &   $<0.357$  			       &$<1.727$\\
CFHT-BL-{\bf 22}  &  00:00:02.661 & -30:20:15.90  &  18.47  &   2.90   &   M5.6   &  $ 24 \pm  5$  &   $-6.22 \pm 0.07$   &   $-4.0863$   &$0.320\,^{+0.137}_{-0.148}$ &$1.387\,^{+0.587}_{-0.566}$\\
CFHT-BL-{\bf 24}  &  00:07:50.539 & -30:05:09.46  &  18.51  &   2.95   &   M6.0   &  $ 3  \pm  8$  &   $-6.47 \pm 0.13$   &   $-4.1053$   &   $<0.291$                 &$<1.374$\\
CFHT-BL-{\bf 25}  &  00:00:02.844 & -30:17:43.98  &  18.62  &   3.06   &   M5.6   &  $ 28 \pm  6$  &   $-6.26 \pm 0.16$   &   $-4.1991$   &   $<0.294$  			   &$<1.313$\\
CFHT-BL-{\bf 29}  &  00:00:17.351 & -30:46:20.32  &  18.77  &   3.03   &   M6.2   &  $ 29 \pm  7$  &   $-5.58 \pm 0.15$   &   $-4.2268$   &   $<0.333$  			   &$<1.490$\\
CFHT-BL-{\bf 38}  &  00:05:13.037 & -30:27:35.78  &  18.98  &   3.10   &   M6.4   &  $ 7  \pm 11$  &   $-4.31 \pm 0.22$   &   $-4.3900$   &$0.957\,^{+0.170}_{-0.145}$ &$2.790\,^{+0.241}_{-0.285}$\\
CFHT-BL-{\bf 43}  &  00:04:32.849 & -30:18:41.40  &  19.02  &   3.13   &   M6.3   &  $ 7  \pm 10$  &   $-6.39 \pm 0.18$   &   $-4.2404$   &$1.183\,^{+0.182}_{-0.122}$ &$2.977\,^{+0.245}_{-0.243}$\\
CFHT-BL-{\bf 36}  &  00:00:28.585 & -30:06:41.94  &  19.06  &   3.37   &   M6.0   &  $ 24 \pm  7$  &   $-5.53 \pm 0.09$   &   $-4.4539$   &   $<0.213$  			   &$<0.827$\\
CFHT-BL-{\bf 45}  &  00:01:35.611 & -30:03:09.90  &  19.23  &   3.27   &   M6.2   &  $ 25 \pm 12$  &   $-5.26 \pm 0.37$   &   $-4.4150$   &$1.521\,^{+0.183}_{-0.117}$ &$3.163\,^{+0.207}_{-0.215}$\\
CFHT-BL-{\bf 49}  &  00:04:28.858 & -30:20:37.00  &  19.46  &   3.56   &   M6.3   &  $ 3  \pm 18$  &   $-2.42 \pm 0.66$   &   $-4.9226$   &$1.930\,^{+0.120}_{-0.166}$ &$3.197\,^{+0.190}_{-0.192}$\\
\enddata
\tablenotetext{a}{Targets are from: B1opt- SMARTS optical survey (James et al. in prep); CFHT-BL- \cite{moraux:2007}.
In this paper, targets will be referenced by the identification number written in bold.}
\tablenotetext{b}{J2000.0 Coordinates}
\tablenotetext{c}{The K$\rm_s$ values come from the 2MASS catalog for stars with $I$$<$17.5, or from \citeauthor{moraux:2007}
for $I$$>$17.5. For $I$$>$17.5, which is relevant to the region of the LDB, the photometric uncertainty is estimated as
$\sigma_{K\rm_s} = 0.03$, $\sigma_{I} = 0.04$, $\sigma_{I-K\rm_s} = 0.05$.}
\tablenotetext{d}{Spectral types are good to within half a subclass.}
\tablenotetext{e}{Negative values indicate the line is in emission.}
\tablenotetext{f}{The systematic uncertainty for $\log L_{\rm H\alpha}/L_{\rm bol}$ is about 0.5 dex.}
\tablenotetext{g}{Members with Li report EW(Li) from our MCMC analysis, while 3$\sigma$ upper limits come from equation \ref{dew}.}
\label{b1}
\end{deluxetable}
\end{landscape}

\clearpage
\subsection{Spectral Indices}\label{spec}

Spectral types are estimated from the value of the TiO (7140\AA) and CaH (6975\AA) narrow-band spectral indices. 
They are defined as
\[ \rm TiO(7140\AA) = \frac{C(7020\!-\!7050\AA)}{TiO(7125\!-\!7155\AA)},\;\; CaH(6975\AA) = \frac{C(7020\!-\!7050\AA)}{CaH(6960\!-\!6990\AA)}, \]
where C(7020--7050\AA) represents the pseudo-continuum, and TiO(7125--7155\AA) and CaH(6960--6990\AA) represent the 
molecular absorption bands, integrated in the indicated wavelength intervals \citep{briceno:1998, oliveira:2003}. 
The CaH index is sensitive to gravity and helps us verify that the objects that we observed are, in fact, dwarfs. 
However, these narrow-band indices are not by themselves good indicators of cluster membership since the sample 
is sure to be contaminated with other foreground field M-dwarfs with similar index values \citep{jeffries:2004}. 
The spectral type of each target is estimated from the relationship between TiO (7140\AA) index and spectral type 
calibrated from standards in \citet{montes:1997fk} and \citet{barrado-y-navascues:1999sj} \citep[see Table~6 in][]{jeffries:2013}. 
The resulting spectral types are reported in Table~\ref{b1}. We adopt a typical uncertainty of half a spectral 
subclass \citep{oliveira:2003,jeffries:2005}.

\begin{figure}[h!]
\centering
 \includegraphics[scale=0.5]{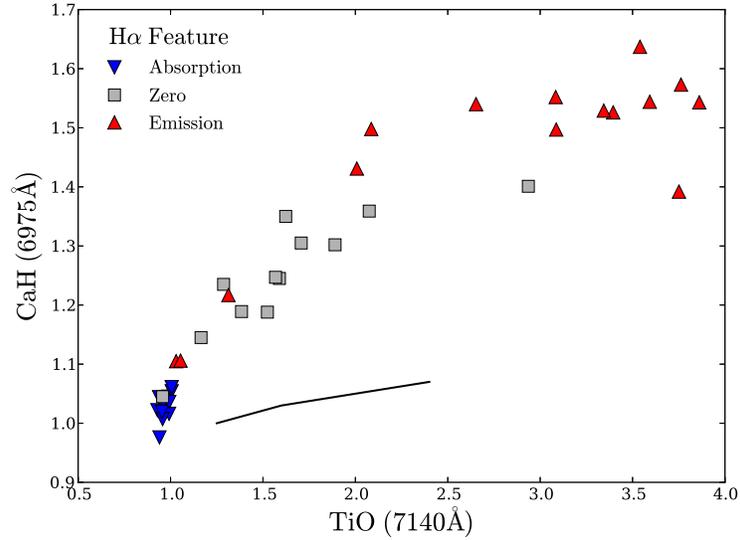}
 \caption{Spectral indices CaH versus TiO for our sample of low-mass Blanco~1 candidates. The spectral type for a
 given object is determined with the TiO index, while the CaH index can be used to eliminate background giant stars 
 from out sample. Note the transition of the \Ha\ feature from absorption to zero to emission as one tends to 
 higher CaH-TiO index (or later M spectral types). The solid line represents the locus of approximate positions
 for giant stars in CaH-vs-TiO space \citep{allen:1995}.}
   \label{f1}
\end{figure}

The CaH versus TiO spectral indices for the Blanco~1 sample are plotted in Figure~\ref{f1}, with reference to an 
H$\alpha$ feature annotated (see \S~\ref{ha}). Stars with \Ha\ in absorption are likely cluster non-members 
because at the age of Blanco~1, we expect such low-mass {\em bona fide} cluster members to be chromospherically active. 
In addition, zero-\Ha\ stars can be very active stars with strong chromospheres as the \Ha\ core may be filled-in 
by active region emission \citep{panagi:1997}; hence, such objects may be young cluster members as well.
We return to \Ha\ as a membership criterion in section~\ref{mem}.

\subsection{The H$\boldsymbol\alpha$ feature}\label{ha}

\begin{figure}[ht]
\centering
 \includegraphics[scale=0.5]{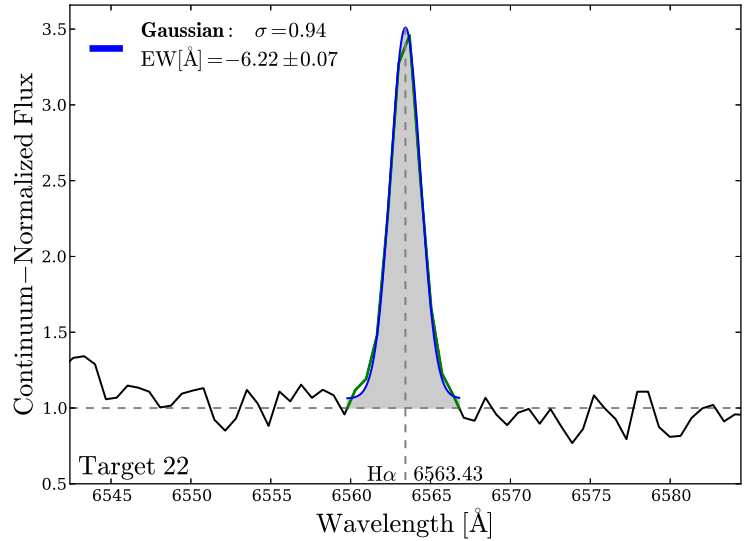}
 \caption{\Ha\ emission from low-mass Blanco~1 member, target 22. The \Ha\ EW is measured over
 the wavelength range where the normalized continuum exceeds unity.}
   \label{22ha}
\end{figure}

As well as establishing cluster membership, \Ha\ EW can be employed in empirical corrections 
for magnetic activity (discussed in section \ref{results}). Our method of obtaining \Ha\ EW 
consistently is achieved by performing continuum normalization with a 10\AA-span of wavelength 
neighboring the \Ha\ feature. We use the intervals 6545--6555\AA\ and 6570--6580\AA\ and 
find no significant issues with other spectral features within these intervals. The mean is 
calculated from each 10\AA-segment, and the line connecting both mean values establishes the 
continuum level. The \Ha\ EW is then determined by measuring excess from the normalized 
continuum using a Gaussian line-profile, which is obtained from a minimized least-squares 
fit. A simple interpolation is performed at the boundaries of the \Ha\ feature so that the 
baseline will exactly measure flux above unity (for emission) or below unity (for absorption). 
Figure~\ref{22ha} demonstates such an EW measurement process for star 22. EW uncertainties 
are estimated from 
\begin{equation}\label{dew}
\rm \sigma_{\rm EW} \simeq 1.5 \times \sqrt{FWHM \times p}/SNR,
\end{equation}
where FWHM, p, and SNR are the full-width half-maximum of the Gaussian fit, the pixel 
dispersion scale in \AA, and the signal-to-noise ratio, respectively \citep{cayrel:1988}.

\clearpage
\subsection{Membership Selection}\label{mem}

The stars from our GMOS sample can be classified as Blanco~1 cluster members
upon consideration of three different membership criteria:
(1) that the photometry of a candidate member is consistent with the cluster sequence in an $I/I-K\rm_s$ CMD;
(2) its 3$\sigma$ RV must be within the range of +2 to +10 km s$^{-1}$;
(3) the \Ha\ line EW must be in emission or zero and comparable to similar-mass stars in the
Pleiades cluster.
We note that Blanco~1 has a measured systemic velocity around +6 km s$^{-1}$
\citep{mermilliod:2008, gonzalez:2009}, but given the difficulty of determining RVs from low SNR spectra,
we consider all stars within +2 to +10 km s$^{-1}$ as candidate cluster members.
Each of these membership criterion has its own level of field star contamination, so each individual property
is considered necessary but not sufficient for cluster membership. We combine the three criteria
so that stars exhibiting these properties are considered high confidence single-star Blanco~1 members.

\begin{figure}[ht]
\centering
 \includegraphics[scale=0.5]{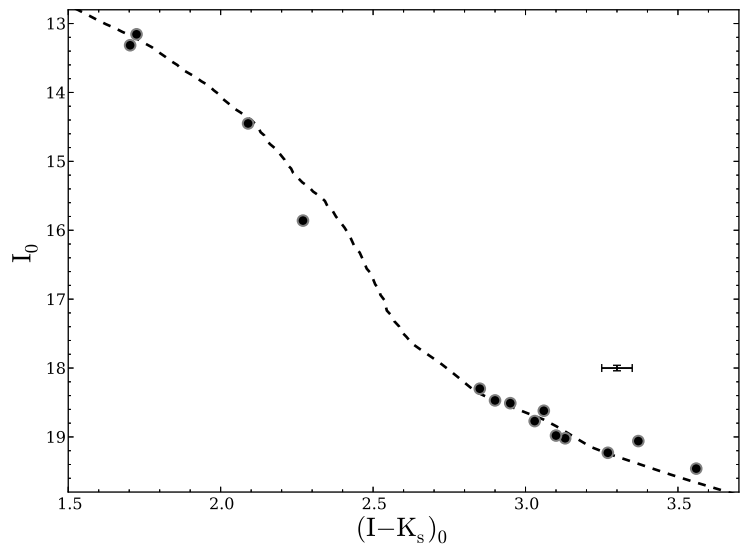}
 \caption{Intrinsic photometry for Blanco~1 low-mass members.
 The Pleiades single-star locus is plotted as the dashed line \citep{stauffer:2007} and is
 shifted appropriately for the distance to Blanco~1 \citep[207 pc,][]{van-leeuwen:2009}.
 The stand-alone error bar represents the estimated uncertainty in photometry.}
   \label{cmd0}
\end{figure}

\begin{figure}[ht]
\centering
 \includegraphics[scale=0.5]{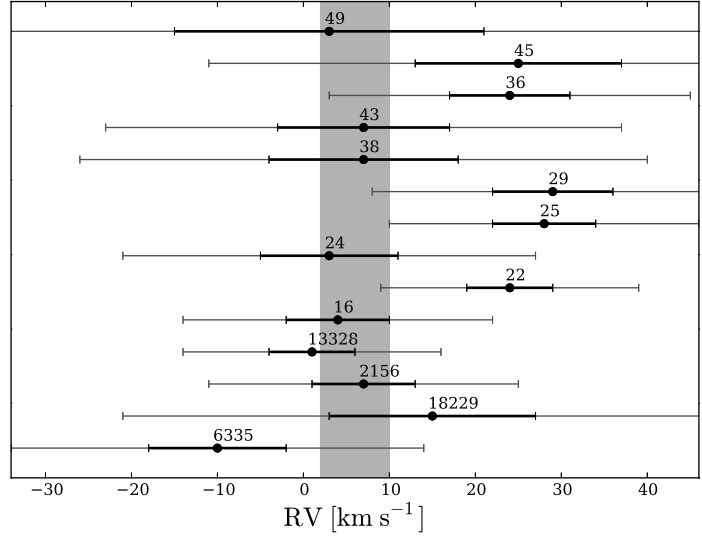}
 \caption{RV distribution for Blanco~1 low-mass members (brighter to fainter going upward).
 The shaded band represents our velocity range for determining RV membership.
 Bold error bars convey 1$\sigma$ errors, while thin error bars show the 3$\sigma$ range.}
  \label{rv}
\end{figure}

\begin{figure}[ht]
\centering
 \includegraphics[scale=0.5]{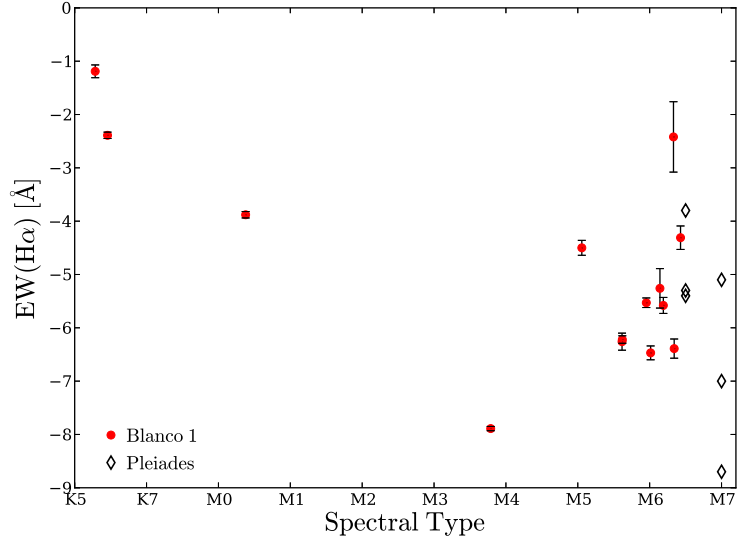}
 \caption{\Ha\ EW versus spectral type for high confidence low-mass
	members of Blanco~1. All of the stars exhibit \Ha\ emission as expected from chromospheric
	activity in young stars. 
	For comparison, we plot several active low-mass Pleiades members from \citet{stauffer:1998aa}.}
   \label{has}
\end{figure}

In Figure~\ref{cmd0}, we plot an intrinsic $I/I-K \rm_s$ cluster sequence for cluster members. We 
correct for reddening and extinction by applying $E(I-K_{\rm s})=0.02$ and $A_I=0.03$ to obtain 
the intrinsic photometry. The RV distribution for Blanco~1 members is shown in Figure~\ref{rv}. 
Most of these low-mass stars fall safely within the range of our RV criterion (shown by the shaded band),
but a few of them appear to be marginal RV members. 
Figure~\ref{has} shows the distribution of EW(\Ha) for our sample of Blanco~1 members and 
some low-mass members in the Pleiades \citep{stauffer:1998aa}. These clusters share a similar 
age, and a given EW(\Ha) is expected to be comparable to similar-mass stars amongst these 
populations. Our sample of low-mass members of Blanco~1 exhibits very active chromospheres
at mid-M spectral types. This appears to be comparable to 
the low-mass activity found in the Pleiades. Recorded in Table~\ref{b1} are our 
measurements of EW(\Ha) for high confidence Blanco~1 members.

\clearpage
\subsection{Lithium}\label{lithium}
\subsubsection{EW Measurement via Spectral Subtraction}\label{liew}

A spectral subtraction technique is carried out in our study by using a catalog of M-dwarf templates 
from the Sloan Digital Sky Survey \citep[SDSS,][]{bochanski:2007}. These templates were produced by 
averaging over 4000 SDSS stellar spectra for spectral types M0--L0. In particular, we use the catalog 
of {\sl inactive} spectra for spectral types M0--M7, where the measured EW of the \Ha\ feature was 
$<$1\AA\ in emission. Moreover, since the majority of the combined spectra used for these templates 
are field M-dwarfs, they are expected to be old enough ($\sim$Gyrs) to have destroyed their initial 
lithium. Due to the lack of K-type templates, we must compare the K-type stars with the M0 SDSS template.
Otherwise, we paired a given GMOS spectrum with a template by rounding to the nearest spectral type
determined from the TiO spectral index described in section \ref{spec}.

\begin{figure}[ht]
\centering
 \includegraphics[scale=0.4]{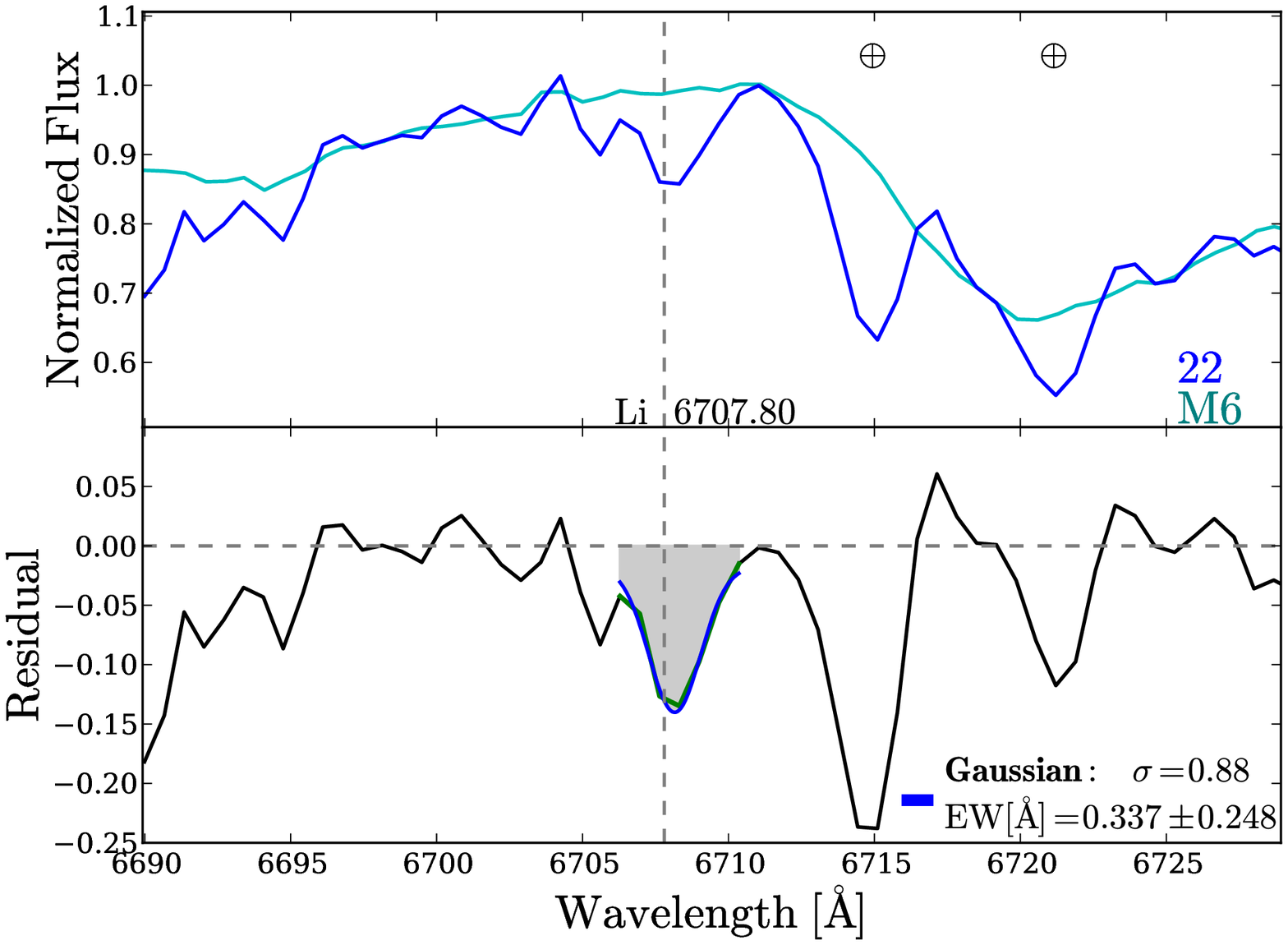}
 \includegraphics[scale=0.4]{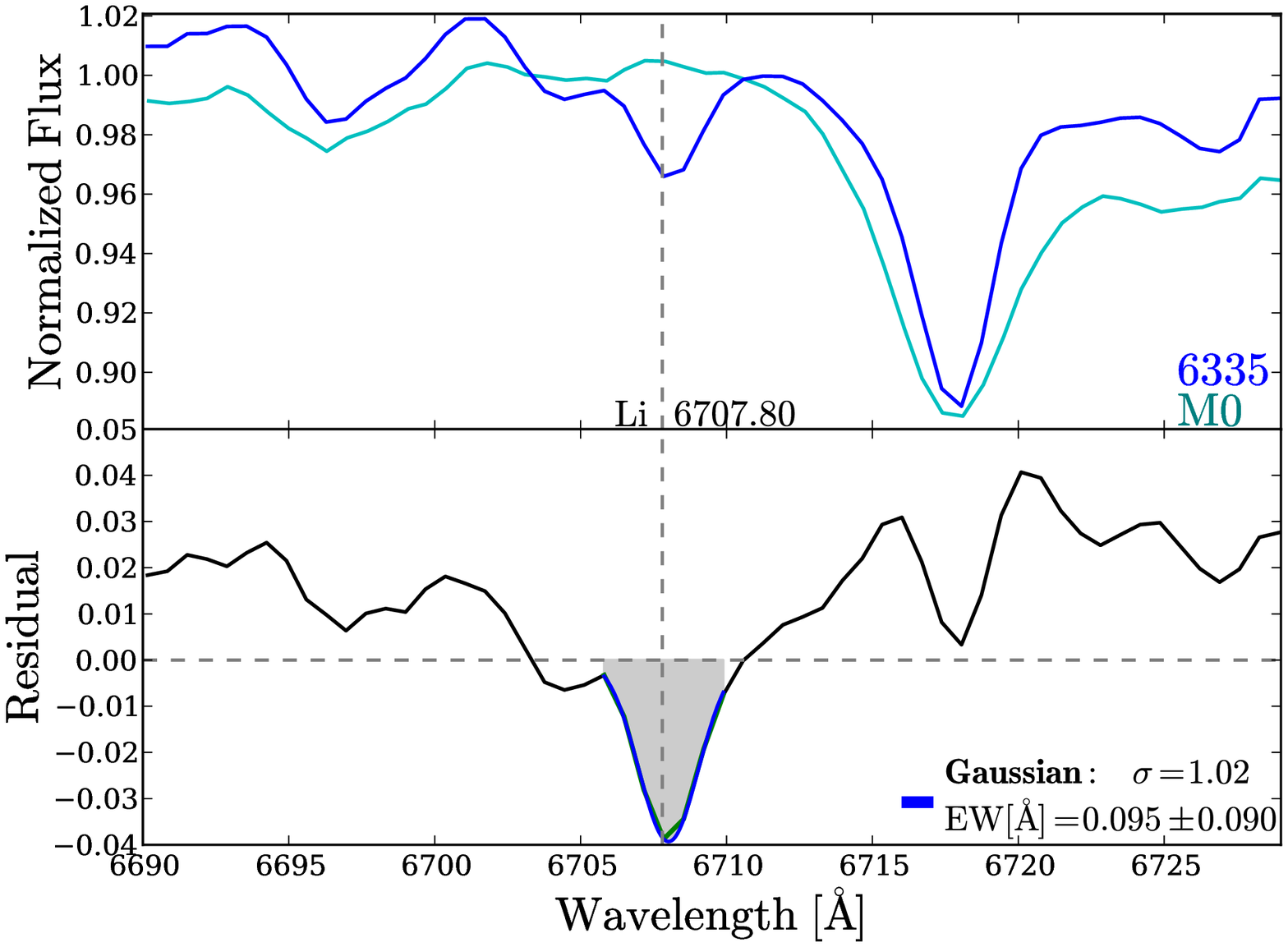}
 \caption{Analysis of the \Li\ feature. Spectral subtraction of SDSS inactive template spectrum
 (teal line) from our GMOS target spectrum (blue line) enables the measurement of EW(Li). The TiO
 index determines which template is most appropriate for subtraction. The measured EW(Li) is
 indicated by the shaded region in the residual. The absorption features denoted
 by Earth symbols are telluric lines near S\,{\sc ii} that could not be subtracted.}
   \label{rli}
\end{figure}

In the determination of the \Li\ EW, the spectrum of the target is shifted to the rest frame, normalized, 
smoothed, and compared with a SDSS template spectrum. Both the target and template are normalized by 
using small wavelength spans bounding \Li\ (6707.8\AA), specifically 6703--6706\AA\ and 6710--6712\AA.
Data were smoothed with a Gaussian kernel, and the template is convolved to match the resolution of our
GMOS spectrum. EWs are measured in the residual spectrum over an interval of $\sim$4\AA\ centered on \Li.
Figure~\ref{rli} shows an example of measuring the Li EW for star 22 (M5.6) and 6335 (K5.5), confirming
detectable lithium in these objects for the first time. Present in some of our spectra with low SNR are 
telluric sky absorption lines near S\,{\sc ii} that could not be removed because of poor sky-subtraction.
The error quoted for EW(Li) in Figure~\ref{rli} is estimated using equation \ref{dew}.

\begin{figure}[ht]
\centering
 \includegraphics[scale=0.5]{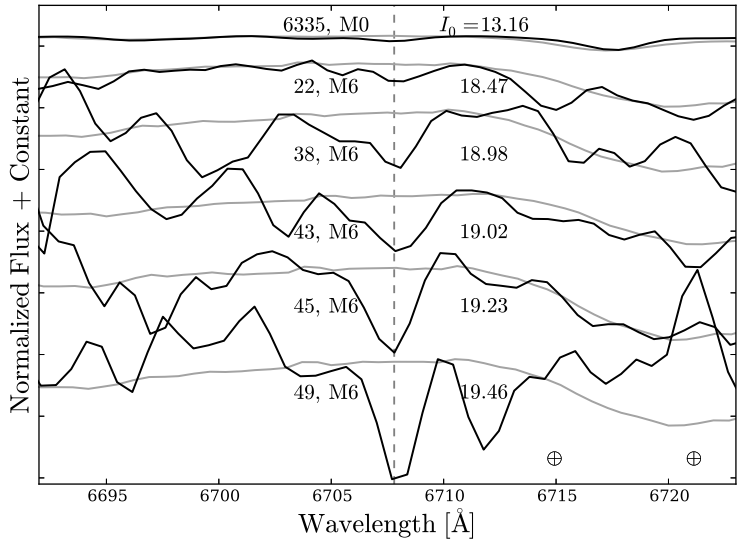}
 \caption{Blanco~1 low-mass members exhibiting \Li\ absorption (grey dashed line). Each Blanco~1 GMOS spectrum
 	is indicated along with its template (solid black and grey lines, respectively) and intrinsic $I$ magnitude.}
   \label{lipanel}
\end{figure}

In figure \ref{lipanel}, we show the six low-mass Blanco~1 members that contain detectable Li.
Telluric S\,{\sc ii} features are indicated.
It is evident that as one goes fainter, the signal in Li becomes more significant.
On the other hand, the SNR diminishes, increasing the difficulty to match the continuum.
Targets 38, 43, 45, and 49 were reported to have Li detected $>$$3\sigma$ in \citet{cargile:2010},
although EW measurements at the time were not possible. Now, we have identified two additional
members (targets 22 and 6335). In section \ref{results}, we explain how target 22 in particular
influences the location of the LDB.

\subsubsection{EW Measurement via MCMC}\label{liMCMC}
Due to the low signal-to-noise of the spectrum around the \Li\ line (typically $\sim$10 for the faintest Blanco~1 stars), we sought to provide a robust characterization of our Li EW measurements. Here, we incorporate an affine-invariant MCMC to sample the posterior probability distribution functions for our \Li\ equivalent widths using the \verb+emcee+\footnote{http://dan.iel.fm/emcee} package developed by \citet{foreman-Mackey:2013}. After we subtract the appropriate template for each target spectrum, we model the resulting residual with a Gaussian as a likelihood function, and calculate 80,000 samples (400 MCMC ``walkers'' $\times$ 200 iteration steps) of the posterior probability distribution. We place an uninformative prior on the amplitude of our Gaussian model, constraining it to only consider Li in absorption, as well as normal priors on the Gaussian $\sigma$ and centroid based on {\em a priori} knowledge of the GMOS instrument resolution and predicted 6707.8\AA\ Li line center, respectively. For each star, we set a conservative estimate of the variance in our flux measurement based on a SNR$=$10.

\begin{figure}[ht]
\centering
 \includegraphics[scale=0.4]{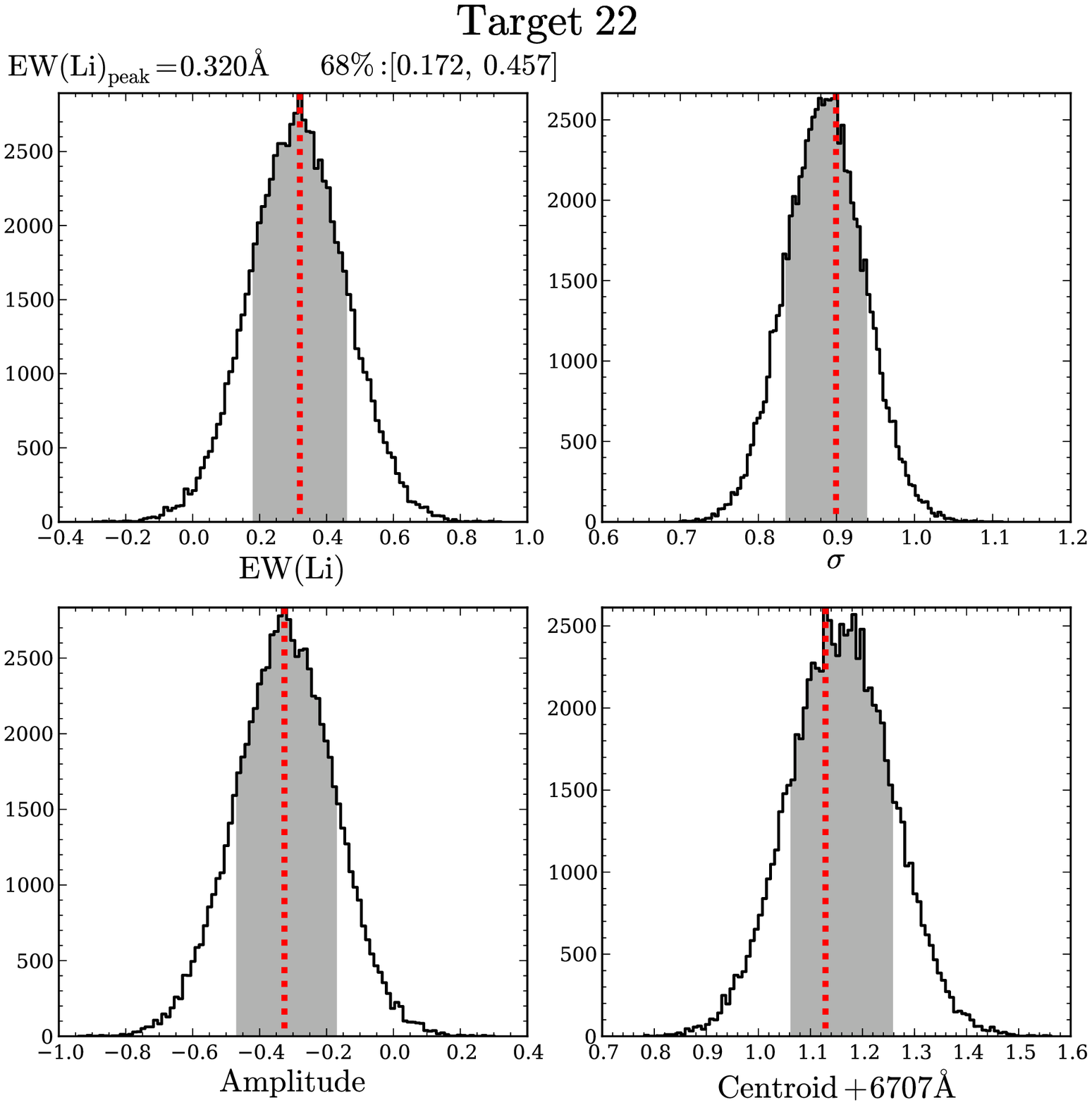}
 \includegraphics[scale=0.4]{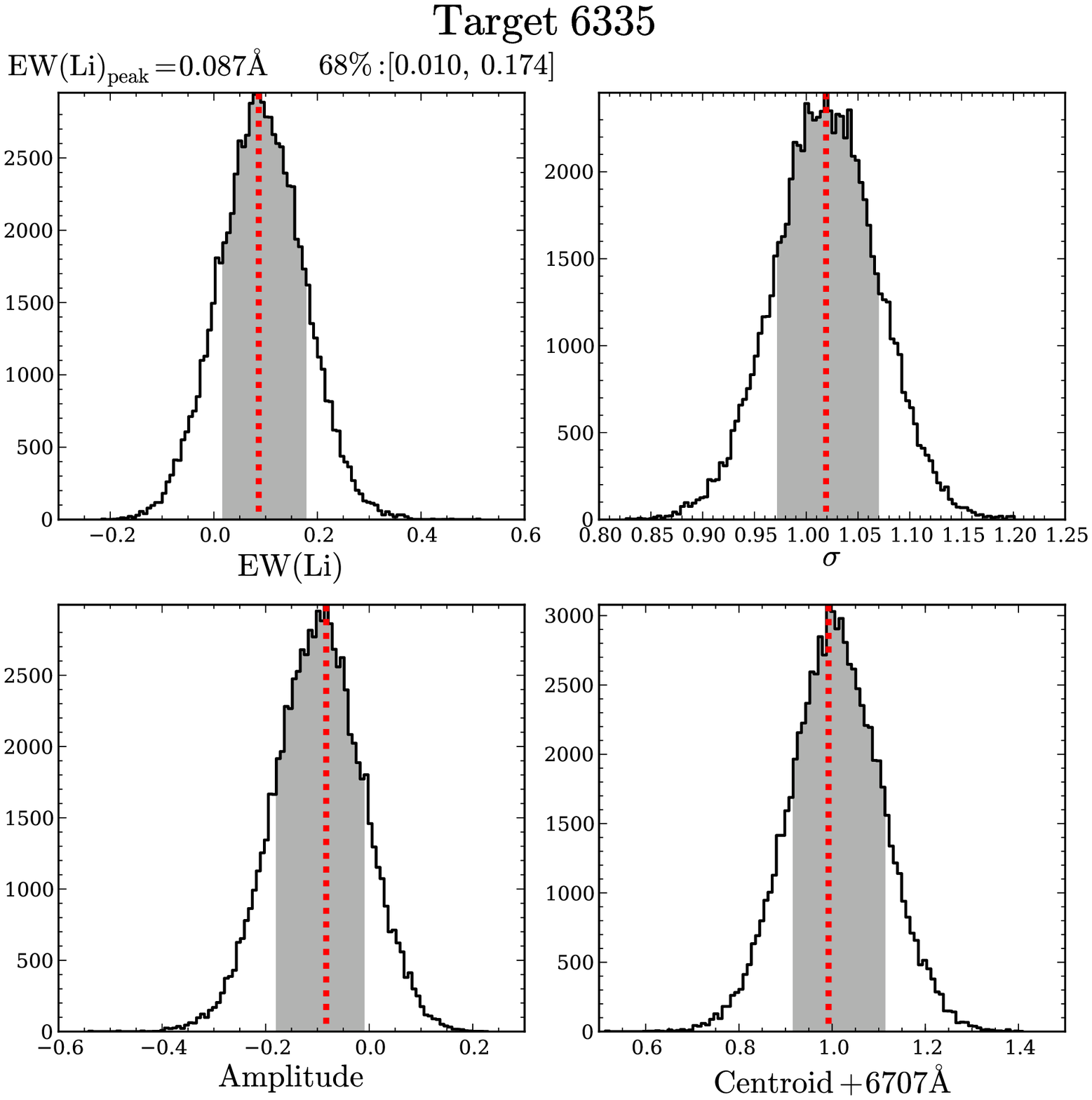}
 \caption{
 Posterior probability distributions for the Gaussian parameters for our modeling of Targets 22 and 6335.
 Best-fit values for the parameters are based on the mode of the distributions (red line) and formal
 68$\rm^{th}$ percentile uncertainty ranges are indicated with shaded regions.
   \label{limcmc}
   }
\end{figure}

We show in Figure \ref{limcmc} examples of the marginalized posterior distributions for a Gaussian model of the Li absorption line, as well as an inferred equivalent width distribution based on the predicted cumulative function. The Li equivalent widths we report in Table~\ref{b1} are determined from the mode of the marginalized distribution with uncertainties based on the inter-68$\rm^{th}$ percentile range (1$\sigma$ errors).

\begin{figure}[ht]
\centering
 \includegraphics[scale=0.5]{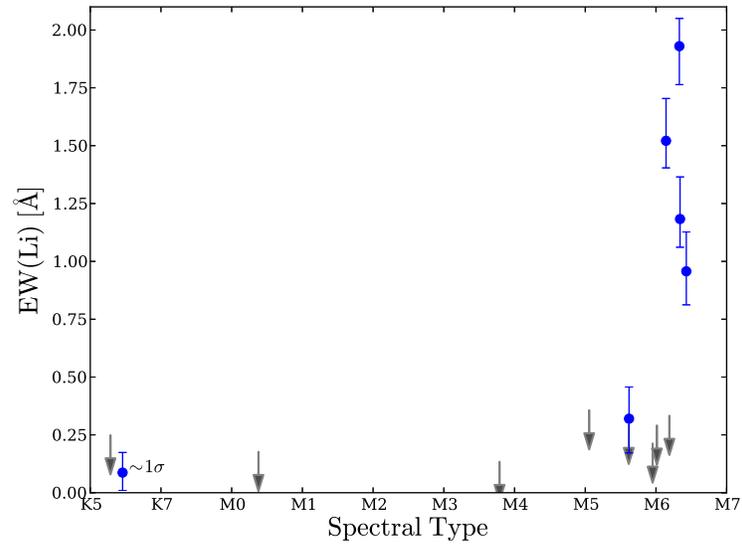}
 \caption{
 Measured \Li\ equivalent width versus spectral type for low-mass members of Blanco~1.
 Stars with detected Li are shown as blue points with 1$\sigma$ errors, and downward arrows
 indicate 3$\sigma$ upper limits for stars with no significant Li absorption based on their GMOS spectra.
 \label{lis}}
\end{figure}

In Figure \ref{lis}, we display the distribution of EW(Li) for our modeling of Blanco~1 cluster members.
A clear pattern is apparent is these data; namely, we detect little or no Li in earlier spectral types ($\lesssim$M6),
but measure significant Li absorption in the latest spectral type stars in Blanco~1. In Section \ref{results},
we further investigate the quantitative nature of this distribution in the context of predictions from PMS Li models.
However, the overall spectral type dependent transition in the EW(Li) of Blanco~1 stars is qualitatively
consistent with the identification of the lithium depletion boundary in the cluster.

\clearpage
\subsubsection{Lithium Abundance}\label{liab}
\begin{figure}[ht]
\centering
 \includegraphics[scale=0.5]{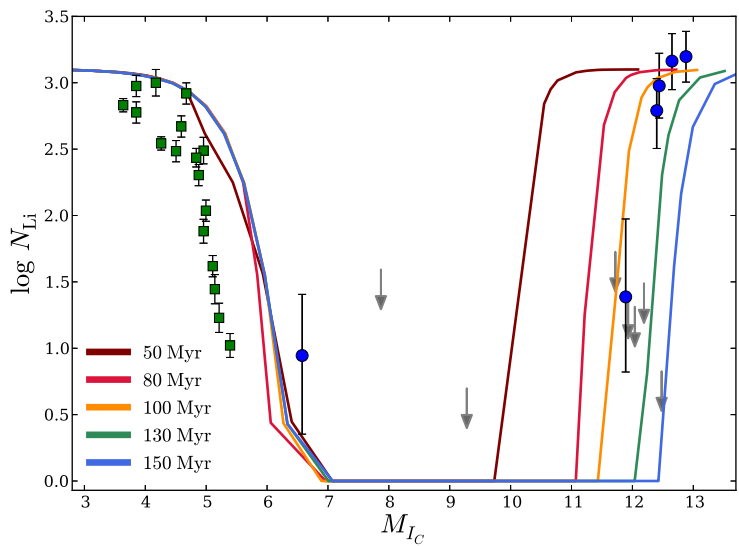}
 \caption{Li abundances for low-mass stars in Blanco~1 shown versus $M_{I_C}$ magnitudes for the
 HIPPARCOS distance modulus to the cluster \citep[6.58 mag,][]{van-leeuwen:2009}. Blue circles represent
 stars in our new sample with Li detections (downward arrows are 3$\sigma$ upper limits), and data from
 \citet[][green squares]{jeffries:1999} show Li abundances for G- and K-dwarf cluster members.
 The Li depletion boundary is located near $M_{I_C}$$\approx$12 with fainter stars retaining their
 full natal Li abundance.}
 \label{abundance}
\end{figure}

In order to calculate Li abundances, it is necessary to convert intrinsic color to $T_{\rm eff}$.
For the stars in our spectroscopic survey, we used BT-Settl models \citep{allard:2011} to obtain $T_{\rm eff}$.
\citet{jeffries:1999} performed a lithium study on G- and K-dwarf members of Blanco~1, but
we use the empirical relationship given in \citet{casagrande:2010} to derive $T_{\rm eff}$ for these stars.
Abundances were calculated from EW(Li) using an appropriate curve of growth for the \Li\ (6707.8\AA)
feature; for hotter stars ($T_{\rm eff}$$>$4000K),
we used the calculations given in \citet{soderblom:1993}, and for cooler objects, we used
the models presented in \citet{pavlenko:1995} and \citet{pavlenko:1996}.
We note that our procedure of measuring EW(Li) after subtracting a template spectrum has
the effect of mitigating the contribution of the nearby contaminating Fe line at 6707\AA,
as well as the large molecular TiO absorption that is present in the Li region.
Non-LTE corrections for Li abundances presented in \citet{carlsson:1994oa}
were applied to the hotter Blanco~1 stars. For the cooler stars, we did not correct
the abundances for non-LTE effects as these are negligible at cool temperatures
\citep{pavlenko:1995,zapatero-osorio:2002}. We adopt an initial Li abundance of
$\log N_{0}({\rm Li}) = 3.1$ for the cluster \citep{zapatero-osorio:2002}.\\
\indent Figure \ref{abundance} shows the distribution of Li abundance for Blanco~1 versus absolute $I$ magnitude,
$M_{I_C}$. Here, the three regimes of Li depletion are present: (1) stars more massive than 0.6 $M_\odot$
gain radiative interiors and only lose a small amount of Li depletion \citep[green squares,][]{jeffries:1999};
(2) stars in the `Li chasm' \citep{basri:1997} that have fully depleted their initial Li supply
($7\lesssim M_{I_C}\lesssim11$);
(3) low-mass stars that still exhibit Li content ($M_{I_C}\gtrsim12$).
It is evident that for hotter stars ($M_{I_C}<6$), with the exception of a few points, the
models fail to reproduce the overall observed abundance distribution in the cluster.
The Li detections near the substellar boundary ($M_{I_C}\approx12$) suggest that target 22 is
currently depleting Li, while the fainter Li detections lie near full natal Li abundance.

\clearpage
\section{Results} \label{results}
\subsection{Locating the LDB} \label{locate}
In the identification of the LDB for this study, we establish a set of rules to demarcate the boundaries
of the LDB. First, having identified the cluster members, we concentrate on the targets that contain lithium.
Based on the insight we have gained from the lithium abundance of Blanco~1 members, the LDB boundaries are
set in the following way:
the target currently depleting lithium (star {\bf 22}) establishes the bright, blue (upper left) corner;
the nearest target in the CMD with full lithium content (star {\bf 38}) establishes the faint, red (lower right) corner.
The edges of the LDB box incorporate the photometric uncertainties in the stars defining these corners (stars 22 and 38):
$\sigma_{K\rm_s}$=0.03, $\sigma_{I}$=0.04, $\sigma_{I-K\rm_s}$=0.05. We define the center of this box to be the location
of the LDB in Blanco~1, the brightest luminosity at which Li content still remains unburned in the
atmospheres of low-mass stars.

\begin{figure}[ht]
\centering
 \includegraphics[scale=0.5]{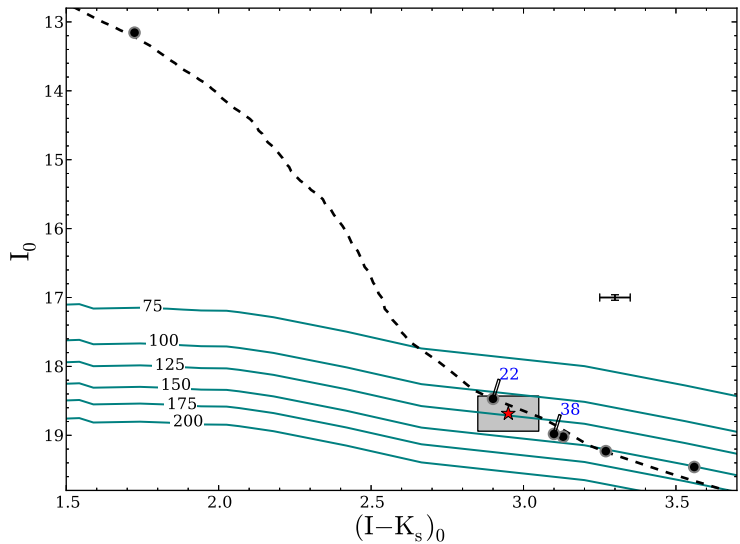}
 \caption{The intrinsic $I/I-K\rm_s$ CMD for the low-mass members of Blanco~1 which exhibit
 Li in their spectra. The shaded rectangle represents the uncertainty on the position of the
 LDB (red star) as established by targets 22 and 38.
 The stand-alone error bar represents the photometric uncertainty.
 The Pleiades single-star locus from \citet{stauffer:2007} is plotted (dashed line), as well as the
 \citet{baraffe:1998aa} predicted luminosity loci for the LDB (solid lines)
 corresponding to the given ages in Myr.}
 \label{cmd}
\end{figure}

Figure \ref{cmd} shows the CMD for intrinsic $I$-band magnitude versus $I-K\rm_s$ for the stars in our sample
that have Li detections among the Blanco~1 low-mass members. The Li detection at $I_0\approx13$ is from a K-dwarf
in the cluster. This star formed a radiative core early enough in its PMS evolution to stop convection down to
the stellar depth necessary to burn Li, and thus still retains some of its Li content. We also illustrate in
Figure \ref{cmd} our definition of the LDB region as the shaded box.

\clearpage
\subsection{LDB Ages for Blanco~1}
\subsubsection{Standard LDB Age}
Previously in \citet{cargile:2010}, with a limited data set, the authors provided a preliminary 
identification of the LDB in Blanco~1 and found it to be located at $I=18.78\pm0.24$ and $I-K\rm_s=3.05\pm0.10$.
They calculated the absolute $I$ magnitude, $M_{I_C}$, of the LDB using the distance modulus from
HIPPARCOS \citep[6.58$\pm$0.12,][]{van-leeuwen:2009} and corrected for reddening and extinction by
adopting $E(I-K\rm_s)=0.02$ and $A_I=0.03$.
Using predicted Li-depletion rates from PMS models, specifically \citet{chabrier:1997}
and \citet[][hereafter BCAH]{baraffe:1998aa}, \citeauthor{cargile:2010} used the luminosity at which 99\% of 
the star's natal Li is destroyed to measure the LDB age for Blanco~1 to be 132$\pm$24 Myr. We designate their 
method as the `standard' LDB age determination technique.

Here, we determine the standard LDB age using a similar approach to \citeauthor{cargile:2010}
Using our stars 22 and 38 to establish the LDB boundaries, the updated Blanco~1 LDB is located
at $I_0 = 18.69\pm0.26$ and $(I-K\rm_s)_0 = 2.95\pm0.10$.
We determine $M_{I_C}$ using the same distance modulus from HIPPARCOS, as well as extinction and reddening 
corrections used in \citeauthor{cargile:2010}
We first calculate the LDB age of Blanco~1 using the BCAH models and synthetic
photometry from the DUSTY model atmospheres \citep{baraffe:2002}. Alternatively, we also calculate
the LDB age using the empirical bolometric corrections from \citet[][hereafter P\&M]{pecaut:2013}
to derive luminosity directly from our absolute $I$ magnitudes.
In Table~\ref{ldbp}, we list our measured LDB parameters for Blanco~1 using the BCAH PMS models with
both synthetic photometry and using empirical corrections from P\&M.

For clarity, Figure \ref{zoomI} shows the region of the $I/I-K\rm_s$ CMD near the LDB of Blanco~1. As in Figure \ref{cmd}, the LDB is established by the Li detections in targets 22 and 38. Using the `standard' LDB technique, the BCAH PMS models with synthetic photometry, and our new Li detections, the LDB in Blanco~1 is found at a $\log(L) = -2.910 L_\odot$, resulting in an updated LDB age of $126^{+13}_{-14}$ Myr. We have included a 126 Myr BCAH LDB luminosity locus in Figure \ref{zoomI} to illustrate this age measurement, which is strikingly similar to the age of the Pleiades \citep[126$\pm$11 Myr,][]{burke:2004vf}.

One might instead consider that the position of the LDB could be defined entirely by Target
22, given that this object evidently lies within the Li depletion zone (see Figure \ref{abundance}).
For stars at full natal Li abundance, A(Li) = 3.1,
so the LDB (defined at 99\% depletion) is found when A(Li) = 1.1. Conceivably, the
absolute $I$ magnitude range for the LDB would be bounded by the abundance errors for star 22.
We interpolate over the abundance isochrones to calculate $M_{I_C}$ for A(Li) = 0.821, 1.1,
and 1.974 dex, which correspond with the error bounds of star 22 and the 99\% Li depletion level.
The 110 Myr model isochrone matches well with the data, and we find at A(Li) = 1.1 that
$dM_{I_C}/d{\rm A(Li)} = 0.214$ mag/dex.
Thus, the LDB using this interpretation is $I_0 = 18.45^{+ 0.19}_{-0.09}$ mag,
and the corresponding LDB age is $114\pm7$ Myr. While this abundance-derived age is in
statistical agreement with our standard LDB age, the reported error (6.1\%) is smaller than
the 10\% systematic error found for the uncertainties associated with the stellar evolution
models and bolometric corrections \citep{burke:2004vf}. Therefore, we prefer the more
conservative approach described above since our method results in the observed precision
of the LDB age that is no better than the predicted accuracy of the LDB technique at $\sim$120 Myr.

\subsubsection{Activity-Corrected LDB Age}\label{activity}
Investigations have found that the fundamental properties of low-mass stars can
be altered in the presence of strong magnetic activity \citep{lopez-morales:2007, ribas:2006}.
\citet{morales:2008} have provided observational evidence that
active stars are cooler than inactive stars of similar
luminosity, therefore, implying that active stars have a larger radius.
Their results generalize for all active low-mass stars -- single or binary.
In the context of the LDB, we thus expect that active stars would be more massive
than initially thought, and their associated ages would be younger.

\citet{stassun:2012} provide empirical relations to determine the amount by which
the effective temperatures ($T_{\rm eff}$) and radii ($R$) of low-mass stars and brown dwarfs
are altered due to chromospheric activity.
Their results presented a strong correlation between the strength of \Ha\ emission in active
M-dwarfs, and the degree to which their temperatures are {\sl suppressed} and
radii {\sl inflated} compared to inactive stars.
In order to determine the change in $T_{\rm eff}$ and $R$ as a result of stellar activity,
the following empirical relations were implemented:
\begin{equation}\label{delt}
\Delta T_{\rm eff}/T_{\rm eff} = m_T \times (\log L_{\rm H\alpha}/L_{\rm bol} + 4) + b_T
\end{equation}
\begin{equation}\label{delr}
\Delta R/R = m_R \times (\log L_{\rm H\alpha}/L_{\rm bol} + 4) + b_R,
\end{equation}
where $m$ and $b$ are linear coefficients. The averaged values, as defined in \citet{stassun:2012}, are in
percent units: $m_T = -4.71 \pm 2.33$, $b_T = -4.4 \pm 0.6$, $m_R = 15.37 \pm 2.91$, and $b_R = 7.1 \pm 0.6$.

We translate our measured \Ha\ EW to $\log L_{\rm H\alpha}/L_{\rm bol}$ using a grid of BT-Settl model
atmospheres from \citet{allard:2011} for $T_{\rm eff}$ in the range 2200--5000K,
assuming Solar composition and $\log(g)=5.0$ (appropriate for very-low-mass stars in Blanco~1).
First, we compute the bolometric flux ($F_{\rm bol}$) for these model atmospheres.
Then, for a given GMOS target, we use its color to estimate $T_{\rm eff}$ from a BCAH 135 Myr isochrone.
This $T_{\rm eff}$ is overestimated since the activity would suppress it, but this is a small effect
($\sim$0.1 dex in $\log L_{\rm H\alpha}/L_{\rm bol}$ for a $\sim$200K shift).
We use this $T_{\rm eff}$ to interpolate over the model atmospheres to estimate the
atmospheric continuum flux at the \Ha\ feature ($F_{\rm \lambda, H\alpha}$).
The \Ha\ flux ($F_{\rm H\alpha}$) is computed by convolving $F_{\rm \lambda, H\alpha}$ with
the \Ha\ EW of our target.
Finally, by computing $\log F_{\rm H\alpha}/F_{\rm bol}$, we also obtain the equivalent $\log L_{\rm H\alpha}/L_{\rm bol}$.
Propagating the photometric uncertainty in color, we find an error of $\sim$0.03 dex in $\log L_{\rm H\alpha}/L_{\rm bol}$,
but this is much smaller than the systematic contribution of $\sim$0.4 dex in the transformation of color to temperature.
Hence, the total systematic error for $\log L_{\rm H\alpha}/L_{\rm bol}$ is about 0.5 dex.

From the empirical relationships, we find the percent change in $T_{\rm eff}$ (suppression)
and $R$ (inflation) as a result of magnetic activity along with the percent change in luminosity.
Due to the nature of how equations \ref{delt} and \ref{delr} were derived and calibrated,
the activity corrections should only be applied to stars with \Ha\ in emission.
We then use this information to determine the 135 Myr BCAH magnitudes and colors of our Blanco~1
sample as if these stars were inactive; we remove the effects of activity.

Using the same logic as before, we set the `corrected' LDB boundaries using the corrected,
inactive photometry for targets 22 and 38.
We determine the LDB parameters at this new LDB location and record these values in Table~\ref{ldbp}.
We infer the activity-corrected LDB is located at $I_0 = 18.45 \pm 0.16$ and $(I-K\rm_s)_0 = 2.77 \pm 0.12$,
which corresponds to $\log(L)=-2.818 L_\odot$ and the BCAH LDB age of $114^{+9}_{-10}$ Myr.

\begin{figure}[ht]
\centering
 \includegraphics[scale=0.6]{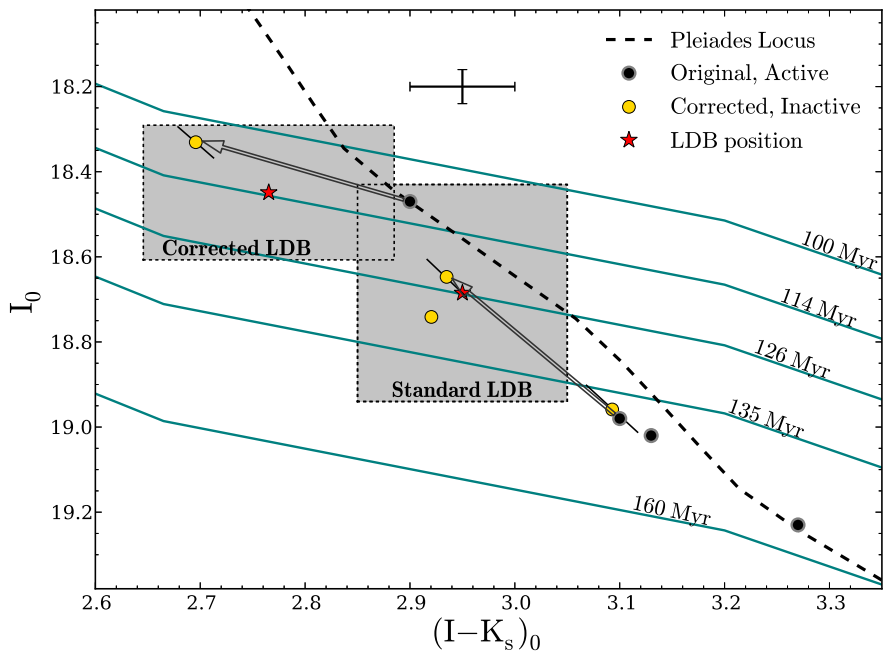}
 \caption{Zoomed in on the regions of the LDB for the Blanco~1 stars with Li detections.
 Stars are shown at their observed CMD position (black points), and our `Standard LDB' location is indicated. 
 Also shown are the shifted locations of the Blanco~1 stars after removing the effects of magnetic 
 activity (yellow points). For simplicity, arrows showing the effect of activity are drawn only for 
 targets 22 and 38. Accounting for stellar activity generally shifts the star's colors/magnitudes 
 brighter and blueward along the cluster sequence. We also show the LDB position after removing the effects of 
 activity on Blanco~1 stars  (`Corrected LDB'). The error bar represents the photometric uncertainty, and the slanted
 error bars on the corrected data show the error due to the uncertainty in the \citeauthor{stassun:2012} empirical relationships.}
 \label{zoomI}
\end{figure}

Figure \ref{zoomI} shows a closer view on the regions of the LDB for the intrinsic
$I/I-K\rm_s$ CMD. This plot shows the Li detections for both the original (black points)
and activity-corrected (yellow points) photometric positions.
For simplicity, vectors showing the direction of the activity corrections are drawn only
for targets 22 and 38, which establish the LDB boundaries in the CMD, and this is shown by
the shaded boxes. The LDB positions (red stars) are marked within these boxes, and BCAH isochrones
are drawn to show the age we infer from their predicted luminosities.
Accounting for the effects of chromospheric activity mainly shifts the data upward
along the cluster sequence.
Additionally, the boundaries of the LDB are compacted when the effects of magnetic
activity are removed and renders a corrected, more precise age.
We carried out this same process of characterizing the LDB for $K\rm_s$ versus $I-K\rm_s$.
LDB parameters derived from PMS models using $K\rm_s$ are recorded in Table~\ref{ldbp}.

Despite consistent treatments, use of the BCAH models along with DUSTY synthetic photometry
renders statistically different LDB ages and parameters depending on whether we use $I_0$ or $K\rm_{s,0}$
magnitudes versus the $(I-K\rm_s)_0$ color as presented in Table~\ref{ldbp}.
For the standard BCAH results, the $I_0$ and $K\rm_{s,0}$ LDB parameter values are statistically
compatible to within 1$\sigma$ of their errors.
Conversely, for the corrected BCAH results, the $I_0$ and $K\rm_{s,0}$ results differ greater than 1$\sigma$.
Further work on understanding the reason for different age and parameter determinations at the LDB 
depending on the choice of photometry is required.

\begin{deluxetable}{lcccc}
\tabletypesize{\footnotesize}
\tablewidth{0pt}
\tablecaption{LDB Parameters for Blanco 1.\vspace{-4mm}}
\tablehead{
 & \colhead{$I_0$ versus $(I-K\rm_s)_0$} & \colhead{$I_0$ versus $(I-K\rm_s)_0$} &
 \colhead{$K\rm_{s,0}$ versus $(I-K\rm_s)_0$} & \colhead{$K\rm_{s,0}$ versus $(I-K\rm_s)_0$}  \\
 & (Standard) & (Corrected) & (Standard) & (Corrected)}
\startdata
{\bf LDB} & $I_0 = 18.685 \pm 0.255$ & $I_0 = 18.450 \pm 0.159$ & $K\rm_{s,0} = 15.685 \pm 0.155$ & $K\rm_{s,0} = 15.635 \pm 0.039$ \\[0.5mm]
{\bf Location}&$(I-K\rm_s)_0 = 2.950 \pm 0.100$ & $(I-K\rm_s)_0 = 2.765 \pm 0.120$ & $(I-K\rm_s)_0 = 2.950 \pm 0.100$ & $(I-K\rm_s)_0 = 2.765 \pm 0.120$ \\[0.5mm]
& $M_{I_C}$\tablenotemark{a} $= 12.114^{+0.280}_{-0.281}$, $^{+0.479}_{-0.575}$ & $M_{I_C} = 11.867\pm0.200$, $^{+0.285}_{-0.347}$
& $M_{K\rm_s} = 12.491^{+0.282}_{-0.290}$ & $M_{K\rm_s} = 12.408^{+0.182}_{-0.184}$\\[1mm]
{\bf Age} [Myr]\\
BCAH\tablenotemark{b} & $126^{+13}_{-14}$ & $114^{+9}_{-10}$ & $145^{+14}_{-15}$ & $141^{+9}_{-10}$ \\
P\&M\tablenotemark{c} & $152^{+24}_{-32}$ & $124^{+14}_{-17}$ & \nodata & \nodata \\[1mm]
$\bm{\log(L)}$ [$L_\odot$]\\
BCAH & $-2.910^{+0.062}_{-0.105}$ & $-2.818^{+0.077}_{-0.075}$ & $-2.993^{+0.066}_{-0.063}$ & $-2.975^{+0.042}_{-0.040}$\\
P\&M & $-3.026^{+0.120}_{-0.108}$ & $-2.898^{+0.080}_{-0.109}$ & \nodata & \nodata \\[1mm]
\enddata
\tablenotetext{a}{Absolute magnitudes are calculated using the HIPPARCOS distance modulus of 6.58 mag.
The first and second sets of uncertainties are when using DUSTY synthetic photometry and the empirical bolometric
corrections of P\&M, respectively. These errors include the uncertainty in the photometry, error in the distance modulus,
and error in the bolometric correction for P\&M. LDB parameters are not available for the $K\rm_s$/$I-K\rm_s$
LDB locations using the P\&M bolometric corrections.}
\tablenotetext{b}{Refers to BCAH models and synthetic photometry from DUSTY model atmospheres \citep{baraffe:2002}.}
\tablenotetext{c}{Refers to use of empirical bolometric corrections from \citet{pecaut:2013} to directly determine luminosity.}
\label{ldbp}
\end{deluxetable}

\clearpage
\section{Summary and Conclusion}\label{summ}
In this paper, we have expanded upon the initial identification of the Blanco~1 LDB \citep{cargile:2010}.
We obtain the full sample of Blanco~1 candidates for our GMOS survey and analyzed both previous and new
data consistently to update the inferred LDB age.
This was done by developing spectral analysis software to systematically analyze the \Ha\ and \Li\ features.
Moreover, we analyze the \Li\ feature using \verb+emcee+ to perform MCMC sampling on the gaussian parameters
that measure EW(Li). We find that for Li detections with $I$$>$17, the error is reduced by up to a factor of 2
using MCMC as opposed to relying on the SNR estimate from equation \ref{dew}.
Since the \Ha\ region is simpler and has a higher SNR, a gaussian fit via least-squares to the \Ha\ line is sufficient.\\
\indent Out of the 43 spectra from our GMOS survey, we find 14 high confidence low-mass members belonging to Blanco~1,
and 6 of these stars exhibit detectable Li features.
Based on our systematic analysis of the \Li\ feature, we verify the findings of \citet{cargile:2010} that targets 
38, 43, 45, and 49 exhibit Li absorption with confidence $>$3$\sigma$.
We have also obtained two new Li detections for low-mass Blanco~1 members; namely, the K-dwarf target 6335 and 
M-dwarf target 22. Importantly, target 22 influences how we determine the LDB age as it appears to currently be
in the process of depleting its initial Li content.\\
\indent Using targets 22 and 38 to establish the LDB boundaries, we derive 
parameters at the LDB using the `standard' technique. 
We first determine the LDB age of Blanco~1 using the BCAH models and synthetic
photometry from the DUSTY model atmospheres, and also obtain the LDB age using the
empirical bolometric corrections from P\&M. Using the BCAH models and the synthetic photometry,
we measure an updated standard LDB age for Blanco~1 of $126^{+13}_{-14}$ Myr.
Compared with the Pleiades \citep[126$\pm$11 Myr,][]{burke:2004vf}, these open clusters share remarkable coevality.\\
\indent For the low-mass Blanco~1 members in our sample, empirical corrections from \citet{stassun:2012}
were used to determine the amount of suppression in $T_{\rm eff}$ and inflation in $R$ due to chromospheric
activity as indicated by \Ha\ emission. We remove these effects and determine the photometric properties
of our targets as if they were not active. Using the inactive properties of targets 22 and 38, we identify
a `corrected' LDB and infer a new age of $114^{+9}_{-10}$ Myr from BCAH models.\\
\indent This corrected age for Blanco~1 brings the LDB age and MSTO isochrone age
\citep[$\tau^2$ isochrone fitting with moderate convective-core overshoot;][]{naylor:2006, naylor:2009}
into close agreement ($\sim$110 Myr, James et al. in prep).
On the other hand, the gyrochronology method from \citet{cargile:2014} determined the age of Blanco~1
to be 146$^{+13}_{-14}$ Myr. Understanding the reasons for this disagreement is beyond the scope of this paper.\\
\indent We find that applying empirical relationships to account for magnetic activity slightly increases the LDB luminosity,
and subsequently results in a $\sim$10\% decrease in the predicted age. This systematic is comparable to the typical
measurement error quoted by other LDB age determinations \citep[\eg][]{burke:2004vf} but has not
been included in any previous LDB study. Our work prompts the need to re-investigate previous LDB determinations
in an effort to produce more accurate ages, and recalibrate the stellar age scale relying on LDB ages.

\acknowledgments
We thank David Soderblom for helpful discussions.
A.J.J. and P.A.C. acknowledge support from the National Science Foundation
Grant AAG Grant AST-1109612.
We gratefully acknowledge the staff at the Cerro Tololo Observatories
and those of the SMARTS Consortium. Our research is
based on observations obtained at the Gemini Observatory,
which is operated by the Association of Universities
for Research in Astronomy, Inc., under a cooperative
agreement with the NSF on behalf of the Gemini
partnership: the National Science Foundation (United States),
the Science and Technology Facilities Council (United Kingdom),
the National Research Council (Canada), CONICYT (Chile),
the Australian Research Council (Australia), Minist\'erio da
Ci\^encia e Tecnologia (Brazil) and Ministerio de Ciencia,
Tecnolog\'ia e Innovaci\'on Productiva (Argentina).

\appendix
\begin{deluxetable}{lccccccccc}
\tabletypesize{\footnotesize}
\tablewidth{0pt}
\tablecaption{GMOS Non-members ordered by Spectral Type. \vspace{-4mm}}
\tablehead{
\colhead{Star ID\tablenotemark{a}} & \colhead{RA\tablenotemark{b}} & \colhead{Dec\tablenotemark{b}} &
\colhead{$I_0$\tablenotemark{c}} & \colhead{$(I-K_{\rm s})_0$\tablenotemark{c}} & \colhead{SpT\tablenotemark{d}} & \colhead{RV} &
\colhead{EW(\Ha)\tablenotemark{e}} & \colhead{EW(\Li)\tablenotemark{f}} \\
\phantom{yo!}&[HH:MM:SS]&[DD:MM:SS]&[mag]&[mag]& & [km s$^{-1}$]&[\AA]&[\AA]}
\startdata
18237 &  00:01:30.043 & -30:02:17.56  &$16.193\pm0.021$&$1.519\pm0.113$&K4.0&$-105 \pm 13$  &$ 1.81 \pm 0.08$   &   $ 0.021 \pm 0.044$\\
222   &  00:08:07.963 & -30:16:55.78  &$11.964\pm0.005$&$1.299\pm0.020$&K4.1&$-156 \pm  5$  &$ 1.29 \pm 0.05$   &   $-0.023 \pm 0.038$\\
3005a &  00:00:16.708 & -30:44:36.56  &  \nodata & \nodata  &   K4.1   &  $ -78 \pm 10$  &   $ 2.36 \pm 0.21$   &   $ 0.110 \pm 0.127$\\
3001a &  00:00:16.656 & -30:47:08.81  &  \nodata & \nodata  &   K4.2   &  $-132 \pm 17$  &   $ 1.43 \pm 0.07$   &   $-0.002 \pm 0.053$\\
3004a &  00:00:17.533 & -30:45:11.62  &  \nodata & \nodata  &   K4.2   &  $-21  \pm  5$  &   $ 1.18 \pm 3.24$   &   $-0.002 \pm 2.506$\\
3006  &  00:00:17.947 & -30:45:20.74  &  \nodata & \nodata  &   K4.2   &  $-10  \pm 13$  &   $ 1.76 \pm 0.07$   &   $-0.014 \pm 0.048$\\
3007  &  00:00:17.856 & -30:48:56.34  &  \nodata & \nodata  &   K4.3   &  $  1 \pm  9$   &   $ 2.13 \pm 0.14$   &   $-0.012 \pm 0.090$\\
2250  &  00:08:11.556 & -30:15:30.27  &  \nodata & \nodata  &   K4.4   &  $ 26 \pm 10$   &   $ 1.37 \pm 0.08$   &   $ 0.008 \pm 0.071$\\
300a  &  00:08:10.980 & -30:17:54.67  &$16.979\pm0.042$&$1.528\pm0.198$&K4.5&$6 \pm 18$  &   $ 2.81 \pm 0.10$   &   $ 0.004 \pm 0.062$\\
3001b &  00:00:29.608 & -30:07:35.47  &  \nodata & \nodata  &   K5.0   &  $-228 \pm 12$  &   $ 2.31 \pm 0.10$   &   $ 0.080 \pm 0.088$\\
2019  &  00:00:03.377 & -30:18:19.04  &  \nodata & \nodata  &   K5.0   &  $-138 \pm  8$  &   $ 1.59 \pm 0.04$   &   $ 0.001 \pm 0.023$\\
2001  &  00:00:02.299 & -30:20:25.55  &  \nodata & \nodata  &   K5.1   &  $-120 \pm 10$  &   $ 2.47 \pm 0.05$   &   $ 0.040 \pm 0.032$\\ 
3004b &  00:00:29.008 & -30:08:07.26  &  \nodata & \nodata  &   K5.1   &  $-129  \pm 99$ &   $ 1.96 \pm 0.13$   &   $ 0.006 \pm 0.227$\\
2002  &  00:00:02.713 & -30:21:41.51  &$16.23$\tablenotemark{g}&$1.635$\tablenotemark{g}&K7.3&$-7  \pm  6$&$ 0.09 \pm 0.70$&$ 0.056 \pm 0.076$\\
3005b &  00:00:28.639 & -30:07:27.73  &$16.807\pm0.04$&$2.189\pm0.05$  &M0.2&$ 20  \pm 5$&   $ 0.03 \pm 0.94$   &   $ 0.062 \pm 0.732$\\
3002a &  00:00:16.786 & -30:48:20.92  &  \nodata & \nodata  &   M0.9   &  $-100 \pm 10$  &   $ 0.17 \pm 0.97$   &   $ 0.000 \pm 0.752$\\
18184 &  00:01:33.739 & -30:06:20.05  &$15.79\pm0.015$&$1.86\pm0.064$  &M2.0&$ 28  \pm  6$&  $ 0.13 \pm 1.14$   &   $-0.062 \pm 0.885$\\
300b  &  00:00:28.379 & -30:09:34.52  &  \nodata & \nodata  &   M2.2   &  $  8  \pm 12$  &   $ 0.68 \pm 1.74$   &   $ 0.062 \pm 1.344$\\
2003  &  00:00:03.452 & -30:19:04.01  &  \nodata & \nodata  &   M2.3   &  $-44  \pm  7$  &   $ 0.06 \pm 0.87$   &   $-0.077 \pm 0.672$\\
3002b &  00:00:28.299 & -30:08:47.11  &  \nodata & \nodata  &   M2.5   &  $-68  \pm 31$  &   $ 0.82 \pm 1.60$   &   $-0.175 \pm 1.239$\\
250   &  00:07:56.902 & -30:04:16.57  &$14.473\pm0.036$&$2.081\pm0.048$&M2.9&$-73  \pm  4$&  $-0.14 \pm 0.82$   &   $-0.124 \pm 0.638$\\
993\tablenotemark{h} &00:05:22.171 & -30:27:59.51&$16.82\pm0.031$&$2.26\pm0.093$&M3.5&$-2  \pm 6$&  $-0.19 \pm 0.86$  &  $ 0.000 \pm 0.161$\\
9424\tablenotemark{h}&00:05:13.306 & -30:26:28.72&$16.37\pm0.022$&$2.26\pm0.068$&M4.0&$ 2  \pm 6$&  $-5.06 \pm 0.06$  &  $-0.019 \pm 0.094$\\
1868\tablenotemark{h}&00:01:36.322 & -30:05:55.39&$17.34\pm0.053$&$2.50\pm0.149$&M4.0&$ 15 \pm 4$&  $-0.21 \pm 0.75$  &  $ 0.029 \pm 0.186$\\
28    &  23:59:55.379 & -30:02:32.28  &$18.75\pm0.04$&$2.80\pm0.05$&     M5.4   &  $-39 \pm  8$  &  $-0.57 \pm 0.69$  &  $ 0.311 \pm 0.535$\\
3     &  23:59:40.898 & -30:01:56.67  &$17.80\pm0.04$&$2.56\pm0.05$&   ---    &   ---   &   ---   &   ---   \\
50    &  23:59:50.002 & -30:01:58.52  &$19.66\pm0.04$&$3.45\pm0.05$&   ---    &   ---   &   ---   &   ---   \\
3003  &  00:00:17.026 & -30:47:43.55  &  \nodata & \nodata  &   ---    &   ---   &   ---   &   ---   \\
9152  &  00:05:26.484 & -30:26:03.77  &$17.357\pm0.051$&$1.864\pm0.228$&   ---   &   ---   &   ---   &   ---   \\
\enddata
\tablenotetext{a}{Targets are from SMARTS optical survey (James et al. in prep). Only targets 28, 3, and 50 are from \cite{moraux:2007}.}
\tablenotetext{b}{J2000.0 Coordinates}
\tablenotetext{c}{The K$\rm_s$ values come from the 2MASS catalog for stars with $I$$<$17.5, or from \citeauthor{moraux:2007}
for $I$$>$17.5.}
\tablenotetext{d}{Spectral types are good to within half a subclass.}
\tablenotetext{e}{Negative values indicate the line is in emission.}
\tablenotetext{f}{Integrative measures of the Li residual. Positive values represent overall absorption. Errors are from equation
\ref{dew}, assuming a FWHM of 1.5\AA.}
\tablenotetext{g}{Not in our optical catalog; $I$ magnitude from USNO-B1.0, $K_{\rm s}$ magnitude from 2MASS.}
\tablenotetext{h}{Proper motion non-members as determined by \citet{platais:2011}.}
\label{nonmem}
\end{deluxetable}

\bibliographystyle{apalike}
\bibliography{LDB-B1}

\end{document}